\documentclass[aps,prd,showpacs,preprintnumbers]{revtex4-2}
\usepackage{amsmath}
\usepackage{graphicx}
\usepackage{float} 
\usepackage{subfigure}

\usepackage{epstopdf}
\usepackage{slashed}
\usepackage{verbatim}

\usepackage{amssymb,amsfonts}
\usepackage{latexsym}
\usepackage{epsfig}
\usepackage{amssymb}
\usepackage{color}
 \usepackage{bm}
 \usepackage{soul}
\newcommand{\be}{\begin{equation}}
\newcommand{\ee}{\end{equation}}

\newcommand{\bea}{\begin{eqnarray}}
\newcommand{\eea}{\end{eqnarray}}

\begin{document}
\title{Magnetism of QCD matter and pion mass from tensor-type spin polarization and anomalous magnetic moment of quarks}
\author{Fan Lin}
\thanks{linfan19@mails.ucas.ac.cn}

\author{Kun Xu }
\thanks{xukun21@ucas.ac.cn}

\author{Mei Huang}
\thanks{huangmei@ucas.ac.cn}
\affiliation{ School of Nuclear Science and Technology, University of Chinese Academy of Sciences, Beijing 100049, China}

\begin{abstract}
We investigate the magnetism of QCD matter and pion mass under magnetic field considering the contribution from the tensor-type spin polarization and the anomalous magnetic moment (AMM) of quarks. It is found that the tensor-type spin polarization (TSP) induces the magnetic catalysis of chiral condensate and diamagnetism (negative magnetic susceptibility) of quark matter at low temperature, both neutral and charged pion masses increase quickly with magnetic field in the case of TSP. The anomalous magnetic moment (AMM) of quarks induces magnetic inhibition and a magnetic dependent AMM causes inverse magnetic catalysis at finite temperature, and the neutral pion mass decreases with magnetic field while the charged pion mass shows nonmonotonic behavior with the magnetic field, which is qualitatively in agreement with lattice result. However, the magnetic susceptibility is positive at low temperature with AMM. In the current framework, our results show the irreconcilable contradiction between the diamagnetism and inverse magnetic catalysis. 
\end{abstract}

\maketitle

\section{Introduction}
Strong magnitude of magnetic field attracts wide interests in early universe, neutron stars and heavy ion collisions. Primordial magnetic fields with magnitude $10^{22}$ ~Gauss can be generated in the early universe driven by chiral anomaly \cite{Joyce:1997uy,Brandenburg:2021aln},  magnitude of $10^{18-19}$ Gauss of magnetic field can be created through non-central heavy-ion collisions in the early stage of quark gluon plasma~\cite{Skokov:2009qp,Deng:2012pc}.  QCD matter under external magnetic fields shows quite a few nontrivial phenomena, for example, Chiral Magnetic Effect(CME)~\cite{Kharzeev:2007tn,Kharzeev:2007jp,Fukushima:2008xe,KharzeevSon:2010gr}, Magnetic Catalysis (MC), i.e., the enhancement of chiral symmetry breaking induced by the magnetic field in the vacuum~\cite{Klevansky:1989vi,Klimenko:1990rh,Gusynin:1995nb} and Inverse Magnetic Catalysis (IMC),i.e., the critical temperature of the chiral phase transition decreases with the magnetic field~\cite{Bali:2011qj,Bali:2012zg,Bali:2013esa}. See reviews in Refs.~\cite{Andersen:2014xxa,Miransky:2015ava,Huang:2015oca,Kharzeev:2015znc,Bzdak:2019pkr} and references therein.

It seems MC in the vacuum will naturally cause the increase of the chiral critical temperature with the magnetic field. However, the lattice results \cite{Bali:2011qj,Bali:2012zg,Bali:2013esa} show the IMC around the critical temperature. There have been many works in the literature trying to explain the IMC, for instance, magnetic inhibition caused by neutral pion fluctuations~\cite{Fukushima:2012kc}, chirality imbalance by sphaleron transition or instanton anti-instanton pairing~\cite{Chao:2013qpa,Yu:2014sla}, and running coupling constant with the magnetic field~\cite{Ferrer:2014qka}. 

It is expected that hadron properties will also change in the presence of magnetic field. Under external magnetic field, the energy of relativistic point particle takes the form of $\varepsilon_{n,s_z}^2(p_z)=p_z^2+(2n-2{\rm sgn}(q)s_z+1)|qB|+m^2$. This gives a linear increase of charged pion mass $m^2_{\pi^{\pm}}(B)=m^2_{\pi^{\pm}}(B=0)+eB$, a constant value of neutral pion and sigma mass under the magnetic field. However, both model \cite{Liu:2014uwa,Liu:2015pna,Liu:2018zag,Wang:2017vtn,Mao:2018dqe,Avancini:2015ady,Avancini:2016fgq,Coppola:2018vkw,Coppola:2019uyr,Fayazbakhsh:2012vr,Fayazbakhsh:2013cha} and lattice calculations show that hadron properties under magnetic field are quite different from the point particle predictions. For example, the MC effect corresponds to the constituent quark mass thus the neutral sigma increasing with magnetic field ~\cite{Klevansky:1989vi,Klimenko:1990rh,Gusynin:1995nb}. Recently, lattice calculations presented some novel properties of pion mass under magnetic field, the mass of neutral pion first decreases then saturates with magnetic field, while the charged pion mass shows a non-monotonic magnetic-field-dependent behavior~\cite{Ding:2020jui}. Furthermore, lattice calculation shows that the magnetized matter exhibits diamagnetism (negative susceptibility) at low temperature and paramagnetism (positive susceptibility) at high temperature~\cite{Bali:2012jv,Bali:2020bcn}.

As it is well-known that, hadrons are not point-like particles but consistent of quarks. Therefore it is natural to consider the effect of quark polarization on hadron spectrum under magnetic field. Calculation in the Namu--Jona-Lasinio (NJL) model taking into account of quark-loop effect, modifies the linear slope of charged pion results on mass spectra~\cite{Liu:2014uwa,Liu:2015pna,Liu:2018zag,Wang:2017vtn,Mao:2018dqe,Avancini:2015ady,Avancini:2016fgq,Coppola:2018vkw,Coppola:2019uyr,Fayazbakhsh:2012vr,Fayazbakhsh:2013cha}.  With considering the quark loop polarization correction in regular NJL model, one can have magnetic catalysis in the vacuum but no IMC at finite temperature. For charged particles, Zeeman splitting is observed for different spin components, the mass spectra are considerably modified by quark loop polarization, but no nonmonotonic behavior for charged pion mass is observed ~\cite{Wang:2017vtn,Mao:2018dqe,Coppola:2018vkw,Coppola:2019uyr,Klevansky:1992qe,He:1997gn,Rehberg:1995nr}. The AMM under magnetic field attracted quite lots of interests recently~\cite{Ferrer:2008dy,Ferrer:2009nq,Ferrer:2013noa,Bicudo:1998qb,Chang:2010hb,Xing:2021kbw,Mao:2018jdo,Mei:2020jzn,Chaudhuri:2019lbw,Ghosh:2020xwp,Strickland:2012vu,Chaudhuri:2020lga,Ferrer:2015wca,Fayazbakhsh:2014mca,Ferrer:2014qka,Chaudhuri:2020lga,Ghosh:2020xwp,Chao:2020jjy}.  The magnetic field catalyzes the chiral condensate, which is composed of pair of quark and antiquark with anti-parallel spins and carries net MM thus triggering dynamical anomalous magnetic moment(AMM) of quark ~\cite{Chang:2010hb,Ferrer:2008dy,Ferrer:2013noa}. 
In Ref.\cite{Xu:2020yag}, by introducing the effect of the anomalous magnetic moments (AMM) of quarks, it was observed the IMC effect at finite temperature, and neutral pion mass decreases very quickly with magnetic field while charged pion mass is not sensitive to the constant AMM. Because of the nonperturbative feature, it is quite challenging to calculate the AMM of quarks under magnetic field. However, the AMM of electron under external magnetic field has been calculated in Ref. \cite{Lin:2021bqv}, and the result shows a magnetic dependent AMM, which motivates us to study the magnetic dependent AMM effect on properties of magnetized quark matter and hadron.  

 The magnetic field also induces spin polarization, i.e., the condensate of quark antiquark pair with parallel spins. As shown in Ref.\cite{Ferrer:2013noa}, a tensor-type interaction $\sim (\bar{\psi}\Sigma^{3}\psi)^2 + (\bar{\psi}i\gamma^5 \Sigma^{3}\psi)^2$ induces a spin-polarization $\langle \bar{\psi}i\gamma^1\gamma^2\psi \rangle$, whose form is similar to the AMM of quarks developed in the presence of magnetic field but charge independent. Comparing with the spin-0 chiral condensate with anti-parallel spin pairing, the parallel spin pairing with spin-1 does not carry net MM.  It is noticed that the tensor polarization operator $\bar{\psi}\sigma^{\mu\nu}\psi$ is also called the SP operator or spin density because $\bar{\psi}\sigma^{12}\psi =\psi^{\dagger}\gamma^0\Sigma^{3}\psi$ with $\Sigma^{3}=\left(\begin{array}{cc} 
\sigma^3~0 \\ 0~\sigma^3\end{array}\right)$  and $\sigma^3=-i \sigma^1\sigma^2$. The SP condensate is for quark anti-quark with parallel spin pairing 
\begin{equation}
\bar{\psi}_f\sigma^{12}\psi_f \sim N_{\uparrow }-N_{\downarrow},
\end{equation}
and the SP condensate in dense quark matter as an origin of magnetic field in compact stars has been investigated in \cite{Tatsumi:1999ab,Maruyama:2000cw,Maruyama:2020kid,Maruyama:2019ter}. 

Here, we compare the contributions to the magnetism of the system from chiral condensate and SP condensate  in a simple constituent quark picture. For a particle with charge $q$, mass $m$ and spin $\vec{s}$, it carries the MM $\vec{\mu}=\frac{q_f}{2m}\vec{s}$. For the chiral condensate with anti-parallel spin pairing of quark anti-quark, it exhibits a net MM, therefore the chiral condensate triggers a dynamical AMM \cite{Bicudo:1998qb,Chang:2010hb,Ferrer:2009nq,Ferrer:2008dy}. Under the presence of an external magnetic field, the net MM tends to align along the direction of the magnetic field.  On the other hand, for SP with parallel spin pairing of quark anti-quark, the MM of the spin aligned quark and anti-quark cancel with each other and the SP pairing does not exhibit net MM. Therefore, the total net MM of the system with SP condensate reduces comparing with that of the system with only chiral condensate carrying net MM. Therefore, it is expected that the system with SP exhibits relative diamagnetism. This senario is similar to the diamagnetic materials where paired electrons in the atoms carrying zero net MM, the difference lies in that there the electron pairs are anti-parallel spin pairing. At high temperature, the quark anti-quark pairs dissociate,  all charged quarks become single small magnets, and turns align along the magnetic field, thus QCD matter at high temperature shows paramagnetism. 


The remaining IMC puzzle and recently discovered properties of magnetized matter attract our renewed interest to revisit QCD vacuum and matter under external magnetic field to find the underlying mechanism for these properties. In this work, We investigate the magnetism of QCD matter and pion mass under magnetic field with the contribution from the tensor-type spin polarization and the anomalous magnetic moment (AMM) of quarks considered, respectively. This paper is organized as follows: in Sec. \ref{sec-model} we introduce two-flavor NJL model with AMM and tensor type interaction in the external magnetic field, and in Sec.\ref{sec-meson}, we investigate the magnetic catalysis and inverse magnetic catalysis by AMM and SP, respectively. Then we investigate the neutral and charged pion masses as functions of magnetic fields with AMM and SP. Finally, we discuss the results in Sec.\ref{sec-con}.

\section{MODEL SETUP}
\label{sec-model}

In this part, we consider two-flavor NJL model including the tensor channel interaction and the AMM of quarks, respectively in the presence of magnetic field.

\subsection{Tensor-type Spin Polarization}

\par Considering the tensor interaction channel into account, the Lagrangian takes the form of  \cite{Ferrer:2013noa}:
\begin{equation}
\mathcal{L}_{\mathrm{T S P}}=\bar{\psi}\left(i \gamma^{\mu} D_{\mu}-m\right) \psi+G_{S}\left[(\bar{\psi} \psi)^{2}+\left(\bar{\psi} i \gamma^{5} \boldsymbol{\tau} \psi\right)^{2}\right]+G_{T}\left[\left(\bar{\psi} \Sigma^{3} \psi\right)^{2}+\left(\bar{\psi} i \gamma^{5} \Sigma^{3} \psi\right)^{2}\right].\label{eq:L1}
\end{equation}
Here $\psi$ is two-flavor quark field $\psi=(u,d)^{\mathbf{T}}$,  $m=m_u=m_d$ is the current mass and we choose the same value for two flavors. The covariant derivative $D_{\mu}=\partial_{\mu}-i q_f A_{\mu}$ with  quarks' electric charge $q_f$, external electromagnetic field $A_{\mu}$ and corresponding field strength tensor $F_{\mu\nu}=\partial_{\mu}A_{\nu}-\partial_{\nu}A_{\mu}$. The term $G_{T}\left[ (\bar{\psi}\Sigma^{3}\psi)^{2}+ (\bar{\psi} i \gamma_{5}\Sigma^{3}\psi)^{2}\right]$ with $\Sigma^{3}=\frac{i}{2}[\gamma^{1},\gamma^{2}]=i\gamma^{1}\gamma^{2}$ will induce the tensor condensate of spin projection.  Besides, $G_S$ and $G_T$ are the coupling constants of (pseudo-)scalar and (axial-)tensor interaction channels, respectively. Under mean-field approximation neglecting the effect from field fluctuation, the Lagrangian is given by:
\begin{equation}
\mathcal{L}_{\mathrm{T S P}}=-\frac{(M-m)^2+\boldsymbol{\pi}^{2}}{4G_S}-\frac{\xi^2+\xi'^{2}}{4G_{T}}+\bar{\psi}(i\gamma^{\mu}D_{\mu}-M-i\xi \gamma^{1}\gamma^{2}-i\xi^{'} \gamma^{0}\gamma^{3})\psi,
\end{equation}
where
\begin{equation}
	\begin{aligned}
		 M= & m+\sigma 
		 \quad \text{with} \quad\sigma=-2 G_{S}\langle\bar{\psi}\psi\rangle,\quad   \boldsymbol{\pi}=-2G_S  \langle\bar{\psi}i\gamma^5\boldsymbol{\tau}\psi\rangle, \\
		 & \xi=-2 G_{T}\langle\bar{\psi}i \gamma^{1}\gamma^{2}\psi\rangle , \quad \xi^{'}=-2 G_{T}\langle\bar{\psi}i \gamma^{0}\gamma^{3}\psi\rangle. 
	\end{aligned}
\end{equation}
\par  
For the tensor channel coupling $G_T$, its magnitude can be derived from Fierz transformation. We can take $ G_T \le G_S $ as a free parameter, a proper coupling choice used in this work is $G_T=G_S/2$. In addition, we drop all pseudo channel condensate, i.e. we assume: $\boldsymbol{\pi}=0$, $\xi^{'}=0$.

\subsection{Anomalous Magnetic Moment Inducing Spin Polarization}

\par Taking the effect of the AMM of quarks into account, the Lagrangian has form of:
\begin{equation}
	\mathcal{L}_{\mathrm{AMM}}=\bar{\psi}\left(i \gamma^{\mu} D_{\mu}-m_{0}+\frac{1}{2} q_f \kappa F_{\mu \nu} \sigma^{\mu \nu}\right) \psi+G_{S}\left[(\bar{\psi} \psi)^{2}+\left(\bar{\psi} i \gamma^{5} \boldsymbol{\tau} \psi\right)^{2}\right],
\end{equation}
 where $\kappa=\mathrm{diag}(\kappa_{u},\kappa_{d})$  and $q_f=\mathrm{diag}(\frac{2}{3},-\frac{1}{3}$) are the AMM and charge matrices of quarks. In above Lagrangian, the effective AMM term $ q_f \kappa F_{\mu \nu} \bar{\psi}\sigma^{\mu \nu}\psi$ originates from the following electromagnetic current\cite{Peskin:1995ev}
\begin{equation}
	\bar{\psi}\Gamma^{\mu}\psi=\bar{\psi}\left[\gamma^{\mu}F_{1}(q^{2})+\frac{\mathrm{i} \sigma^{\mu\nu} q_{\nu}}{2m} F_{2}(q^{2})\right]\psi. \label{EMC}
\end{equation}

Obviously, this term is invariant under $U(1)$ gauge transformation but variant under global chiral transformation $\psi \rightarrow \exp(\mathrm{i}\theta\gamma^{5})$, which indicates $F_{2} \equiv 0$ for massless fermion in the Wigner mode with chiral symmetry. If one only considers the QED vacuum accompanied by slightly broken chiral symmetry, the vacuum expectation of the operator $\bar{\psi}\sigma^{\mu\nu}\psi$  is too small to affect the interaction of QCD scale, especially the AMM of electron $ F_{2}(0) \sim 10^{-3}$ does not matter. However, the case is much more complicated for QCD vacuum because the dynamical chiral symmetry breaking causes considerable quark condensate.  Based on constituent quark scenario, the proton and neutron magnetic moment gives $ \kappa_{u} \sim 0.290 \mathrm{GeV}^{-1} , \kappa_{d} \sim 0.360\mathrm{GeV}^{-1}$\cite{Fayazbakhsh:2014mca}. By using nonperturbative Dyson-Schwinger equation, L.Chang et al. have argued that the effective quark mass from dynamical chiral symmetry breaking (DCSB) naturally enter electromagnetic current Eq.(\ref{EMC}). Thus dynamical mass $m_{\mathrm{dyn}}$ from DCSB contributes to AMM $\kappa  \varpropto m_{\mathrm{dyn}} $, so we assume $ \kappa_{u}=\kappa_{d}=\kappa= 0.75 \,\sigma$ for quarks (neglect small AMM of current quark) which in the vacuum is in agreement with the magnitude of  
$ \kappa_{u} \sim 0.290 \mathrm{GeV}^{-1} , \kappa_{d} \sim 0.360 \mathrm{GeV}^{-1}$. Once the temperature increases and exceeds the critical point $T_{c}$, $ \kappa $ decreases and turns to zero implying dynamical AMM disappears while the chiral symmetry restores. In \cite{Lin:2021bqv}, the AMM of electron under magnetic field has been calculated and it was shown that it is proportional to the $m^2$, therefore, we can extend it to QCD as $\kappa= v \sigma^2$ with $v= 2.30$.

\subsection{Dispersion Relation of quarks}
\label{app_DIS}
 Now, we derive the dispersion relation for quark spinor $\psi$ with charge $q_f$, dynamical mass $M$ and TSP condensate $\xi$ in the presence of homogeneous magnetic field $B$ where the Dirac equation is given by
\begin{equation}
\label{equ:Dirac1}
(i\gamma^{\mu}D_{\!\mu}-M-i\xi\gamma^{1}\gamma^{2})\psi=0.
\end{equation}
Same as in Eq.(\ref{eq:L1}), we assume that the magnetic field orients to $z$ direction and choose a gauge $A^{\mu}=(0,0,Bx,0)$, corresponding to $\mathbf{B}=\boldsymbol{\nabla} \times \mathbf{A}=(0,0,B)$. In the following we always assume $ B > 0 $ for convenience. In this work, we use the Ritus method \cite{Ritus:1972ky,Fukushima:2018grm} to do calculations. Constructing the $4 \times 4$ projection matrix
\begin{equation}
P_{n, p}(x) \equiv \frac{1}{2}\left[f_{n, p}^{+}(x)+f_{n, p}^{-}(x)\right]+\frac{\mathrm{i}}{2}\left[f_{n, p}^{+}(x)-f_{n, p}^{-}(x)\right] \gamma^{1} \gamma^{2},
\end{equation}
where the harmonic oscillator wave-functions have the form of
\begin{equation}
\begin{aligned}
f_{n, p}^{+}(x) & \equiv \mathrm{e}^{-\mathrm{i}\left(\omega t-p_{y} y-p_{z} z\right)} \phi_{n}\left(x-p_{y} / |q_f| B\right), & &(n=0,1, \ldots) \\
f_{n, p}^{-}(x) & \equiv \mathrm{e}^{-\mathrm{i}\left(\omega t-p_{y} y-p_{z} z\right)} \phi_{n-1}\left(x-p_{y} /|q_f| B\right), & &(n=1,2, \ldots)
\end{aligned}
\end{equation}
with $\phi_{n}(x)$, the wave-function of the harmonic oscillator relevant to the magnetic magnitude:
\begin{equation}
\phi_{n}(x) \equiv \sqrt{\frac{1}{2^{n} n !}}\left(\frac{ |q_f|  B}{\pi}\right)^{1 / 4} e^{-\frac{1}{2} |q_f| B x^{2}} H_{n}(\sqrt{|q_f|  B} x),
\end{equation}
here $H_n(x)$ represents the Hermite polynomials.
With the projection matrix and property of special function $H_n(x)$ we can easily find
\begin{equation}
(i \slashed{\partial}-q_f  \slashed{A}-M) P_{n, p}(x)=P_{n, p}(x)\left(\omega \gamma^{0}+\sqrt{2 n  |q_f|  B } \gamma^{2}-p_{z} \gamma^{3}-M\right),
\end{equation}
$P_{n, p}(x)$ plays as an eigenfunction of operator $i \slashed{\partial}-q_f  \slashed{A}-M$. From the  relation $ \gamma^{1}\gamma^{2} P_{n, p}(x)=P_{n, p}(x)\gamma^{1}\gamma^{2} $, one  can obtain
\begin{equation}
(i \slashed{\partial}-q_f  \slashed{A}-M-i\xi\gamma^{1}\gamma^{2})P_{n, p}(x)=P_{n, p}(x)\left(\omega \gamma^{0}+\sqrt{2 n  |q_f|  B } \gamma^{2}-p_{z} \gamma^{3}-M-i\xi\gamma^{1}\gamma^{2}\right).
\end{equation}
Therefore the solutions of the Dirac equation under constant $\mathbf{B}$ decompose into 
\begin{equation}
\psi(x)= P_{n, p}(x) u(\tilde{p}) \quad \psi(x)= P_{n, p}^{*}(x) v(\tilde{p})
\end{equation}
for particle  and anti-particle separately, the $u(\tilde{p}),v(\tilde{p})$ are Dirac spinors with momentum $ \tilde{p}=\left(\omega, 0,-\sqrt{2 n  |q_f|  B }, p_{z}\right)$ which means
\begin{equation}
\left(\slashed{\tilde{p}}-M-i\xi\gamma^{1}\gamma^{2}\right)u(\tilde{p}) / v(\tilde{p})=0.
\end{equation}
To derive the dispersion relation, the determinant of coefficient matrix about spinors should be zero
\begin{equation}
\operatorname{Det}[\slashed{\tilde{p}}-M-i\xi\gamma^{1}\gamma^{2}]=\left[(M^{2}-(|\tilde{p}_{\parallel}|+\xi)^{2}+|\tilde{p}_{\perp}|^{2}\right]\times\left[(M^{2}-(|\tilde{p}_{\parallel}|-\xi)^{2}+|\tilde{p}_{\perp}|^{2}\right]=0,
\end{equation}
with new notations $\tilde{p}_\parallel=\left(\omega,0,0,p_z\right)$ and $\tilde{p}_\perp=\left(0,0,-\sqrt{2 n  |q_f|B},0\right)$ introduced. 
After some simple calculations the dispersion relation reads 
\begin{equation}
	\label{eq:dispersion}
E_{f,n,s}^2 = 
\begin{cases} 
p_z^2+\left(\sqrt{M_f^2+2n |q_f| B}-s \xi \right)^2, & n \ge 1 .\\
&   \\
p_z^2+\left( M_f+\xi \right)^2, &  n=0 .
\end{cases}
\end{equation}
Where $ M_f=\sigma+m_f $ and  $s=\pm 1$ correspond to different spin projections. Especially, the quark's spin in lowest Landau level ($ n=0 $) has definite spin projection. For higher excitations ($ n \ge 1 $), the energy spectrum with TSP condensate $\xi$ exhibits Zeeman splittings ($s=\pm1$) .

\par The operation for the Dirac equation with the AMM  of quarks is similar:
\begin{equation}
	(i \gamma^{\mu} D_{\mu}-M+\frac{1}{2} \kappa q \sigma^{\mu \nu} F_{\mu \nu}) \psi=0,
\end{equation}
one just needs to replace $ \xi $ with $ \kappa q_f B $ and replace the Landau level $ 2n= 2l+1-s\zeta , \zeta=\mathrm{sign}\left(q_f\right), k=0,1,2,\dots, s=\pm 1$, it is not difficult to write down the dispersion relation of quarks with AMM  \cite{Xu:2020yag}
\begin{equation}
	\omega_{f,l,s}^{2}=\left\{
	\begin{array}{ll}
		p_{z}^{2}+\left[\sqrt{M_f^{2}+\left(2l+1-s\zeta\right)|q_f B|}-s\kappa q_f B\right]^{2}, & l \geq 1 .\\
		&   \\\label{key}
		p_{z}^{2}+\left(M_f-\kappa |q_f|B\right)^{2}, &  l=0 .
	\end{array}
\right.
\end{equation}
As is shown, the AMM $\kappa$ induces a charge-dependent tensor condensate $\langle\bar{\psi}\sigma^{\mu\nu}\psi\rangle=-\kappa q_{f}F^{\mu\nu} $ whose non-zero components are $\langle\bar{\psi}\sigma^{12}\psi\rangle,\langle\bar{\psi}\sigma^{21}\psi\rangle$ in uniform magnetic along $z$ direction.

\par In qualitative view, TSP is charge independent and elevates the lowest Landau level but the effect of AMM is absolutely opposite, which suggests the TSP and AMM express MC and IMC, respectively.

\subsection{Thermodynamic potential and gap equations}
\label{app_gap}
From the dispersion relation Eq.(\ref{eq:dispersion}), the TSP condensate increases or decreases the dynamical quark mass. The ground state should be determined by the thermodynamic potential.  Following similar procedures like \cite{Ferrer:2013noa,Buballa:2003qv,Chaudhuri:2019lbw}, we obtain the thermal potential one-loop level with finite temperature:
\begin{equation}
\label{equ:gapequT}
      \Omega_{\mathrm{TSP}}=\frac{\sigma^2}{4G_S}+\frac{\xi^2}{4G_T}  - N_c\sum_{f,n,s} \frac{|q_{f}B|}{2\pi}\int_{-\infty}^{+\infty} \frac{\mathrm{d}p_z}{2\pi}E_{f,n,s} -N_c\sum_{f,n,s}\frac{|q_{f}B|}{2\pi}\int_{-\infty}^{+\infty} \frac{\mathrm{d}p_z}{2\pi} 2T \ln(1+\text{e}^{-\beta E_{f,n,s}}).
\end{equation}
where $\beta=1/T$ and the summation is taken over all Landau levels. The first two terms stem from two types four-vertex interactions: scalar and tensor channels, respectively. The last two terms originate from quasi-particles with dispersion relation in Eq.(\ref{eq:dispersion}) . 

\par The minimum  of the thermodynamic potential determines the chiral condensate and tensor-type spin polarization, which is equivalent to solve the gap equations
\begin{equation}
	\frac{\partial \Omega_{\mathrm{TSP}}(\sigma,\xi)}{\partial \sigma }  =0 \ , \quad \frac{\partial \Omega_{\mathrm{TSP}}(\sigma,\xi)}{\partial \xi }  =0 \ , \qquad \sigma\ > 0 ,\ \xi > 0.
\end{equation}
Explicitly, we have the two coupled gap equations:
\begin{equation}
\begin{aligned}
\frac{\sigma}{2 G_S}=&N_c\sum_{f,n,s} \frac{|q_{f}B|}{2\pi}\int_{-\infty}^{+\infty} \frac{\mathrm{d}p_z}{2\pi} \left[ 1-2(1+\text{e}^{-\beta E_{f,n,s}})^{-1}\right]\frac{M_{f,0} }{E_{f,n,s}}(1-\frac{s\xi}{M_{f,n}}),\\
\frac{\xi}{2 G_T}=&N_c\sum_{f,n,s} \frac{|q_{f}B|}{2\pi}\int_{-\infty}^{+\infty} \frac{\mathrm{d}p_z}{2\pi} \left[1-2(1+\text{e}^{-\beta E_{f,n,s}})^{-1} \right] \frac{1}{E_{f,n,s}}(\xi-sM_{f,n}).
\label{gapeq-xi-chiral}
\end{aligned}
\end{equation}
Where $M_{f,0}=M_f=m_f+\sigma$ is dynamical mass of quark and $M_{f,n}=\sqrt{M_f^2+2n|q_fB|}$.

\par Similarly, the thermodynamic potential with AMM as well as its corresponding gap equation under the mean-field approximation read:
\begin{equation}
	\Omega_{\mathrm{AMM}}=\frac{\sigma^2}{4G_S} - N_c\sum_{f,l,s}\frac{|q_{f}B|}{2\pi}\int_{-\infty}^{+\infty} \frac{\mathrm{d}p_z}{2\pi}\omega_{f,l,s} -N_c\sum_{f,l,s}\frac{|q_{f}B|}{2\pi}\int_{-\infty}^{+\infty} \frac{\mathrm{d}p_z}{2\pi}\ 2T \ln(1+\text{e}^{-\beta \omega_{f,l,s}}),
\end{equation}
\begin{equation}
	\frac{\partial \Omega_{\mathrm{AMM}}}{\partial \sigma } =\frac{\sigma}{2G_{s}} - N_c\sum_{f,l,s}\frac{|q_{f}B|}{2\pi}\int_{-\infty}^{+\infty} \frac{\mathrm{d}p_z}{2\pi} \left[ 1-2(1+\text{e}^{\beta \omega_{f,l,s}})^{-1}\right]\frac{\partial\omega_{f,l,s}}{\partial \sigma}=0, \quad \sigma>0. 
\end{equation}

 The form of the AMM of quarks under the magnetic field is not known, we assume it can take the nonperturbative form of $\kappa=v \sigma$ or $\kappa=v \sigma^2$. If we choose $\kappa= v\sigma^{2}$ and use LLL approximation for a qualitative analysis at strong magnetic field, then the gap equation has the following expression:
\begin{equation}
	\frac{\sigma}{2 G_S}=N_c\sum_{f} \frac{|q_{f}B|}{2\pi}\int_{-\infty}^{+\infty} \frac{\mathrm{d}p_z}{2\pi}\frac{(1- 2 v\sigma |q_{f}|B)(\sigma+m_{f}-v \sigma^{2}|q_{f}|B)}{\sqrt{p_z^2+\left( \sigma+m_{f}-v \sigma^{2}|q_{f}|B \right)^2}} ,
	\label{gapeq-AMM-chiral}
\end{equation}

 \section{Numerical results}
 
\par The NJL model with four-fermion interaction is non-renormalized, thus a reasonable cut-off is necessary to avoid ultraviolet divergence. The soft cut-off applied here in $z$ direction for vacuum contribution part of thermodynamic potential is dependent on Landau levels :
\begin{equation}
\frac{|q_f B|}{2\pi}\sum_{n}\int_{-\infty}^{+\infty} \frac{\mathrm{d}p_z}{2\pi} \quad\longrightarrow\quad \frac{|q_f B|}{2\pi}\sum_{n}\int_{-\infty}^{+\infty} \frac{\mathrm{d}p_z}{2\pi}f_\Lambda (p_z,n)
\end{equation}
with
\begin{equation}
\label{equ:cutofffun}
f_\Lambda (p_z,n)=\frac{\Lambda^{10}}{\Lambda^{10}+(p_z^2+2n|q_f B|)^{5}}.
\end{equation}

\par In this paper, we choose two-flavor current quark mass $m_u=m_d=m=5.0\,\mathrm{MeV}$ . The momentum cutoff parameter $\Lambda$ in Eq.(\ref{equ:cutofffun}), (pseudo-)scalar/(pseudo-)vector coupling constant $G_S$ and $G_V$ are determined by fitting to experimental data at zero temperature and vanishing magnetic field. In this work, we use the parameters set : $\Lambda=624.18\text{MeV}$ and $G_S \Lambda^2=2.014$ fitted with pion decay constant $f_{\pi}=93\text{MeV}$, pion mass $m_{\pi}=135.6\text{MeV}$ as well as the quark condensate $\langle\bar{\psi} \psi \rangle=-(251.8\text{MeV})^3$.
\subsection{Magnetic catalysis and Magnetic inhibition with AMM and TSP}

\par \textbf{Magnetic Catalysis with TSP:} For tensor-type spin polarization, in the case of $ eB \sim \Lambda^2$, and the magnetic field magnitude is large enough that LLL dominates the infrared dynamics, which greatly simplifies the gap equations at zero temperature
\begin{equation}
 \frac{\sigma}{G_S}=\frac{\xi}{G_T}, \quad  \frac{\sigma}{2 G_S}=N_c\sum_{f} \frac{|q_{f}B|}{2\pi}\int_{-\infty}^{+\infty} \frac{\mathrm{d}p_z}{2\pi}\frac{M_f+\xi}{\sqrt{p_z^2+\left( M_f+\xi \right)^2}} .
\end{equation}
Neglecting the small current quark mass temporally and using the hard cut-off $ \Lambda $, the analytical solution can be found 
\begin{equation}
	\sigma = \left(\frac{G_S}{G_S+G_T}\right) \frac{\Lambda}{\sinh\!\zeta},  \quad 	\xi = \left(\frac{G_T}{G_S+G_T}\right) \frac{\Lambda}{\sinh\!\zeta}, \quad \text{with} \quad 
	\zeta = \frac{\pi^2}{(G_S+G_T)N_c\sum\limits_{f} |q_{f}B|}.
\end{equation}
Applying the asymptotic expansion  $\sinh\!\zeta = \zeta +\zeta^{3}/6+ \mathcal{O}(\zeta)$, we can see that the TSP $\xi$ increases with the magnetic field, and the quark condensate $\sigma $ or quark dynamical mass $ M_f= m_f +\sigma $ increases slightly compared with the case of  $\xi=0$, which indicates that the TSP enhances the magnetic catalysis slightly. 

\begin{figure}[H]
 	\centering 
 	\subfigure[]{
 		\label{Fig1a}
 		\includegraphics[scale=0.55]{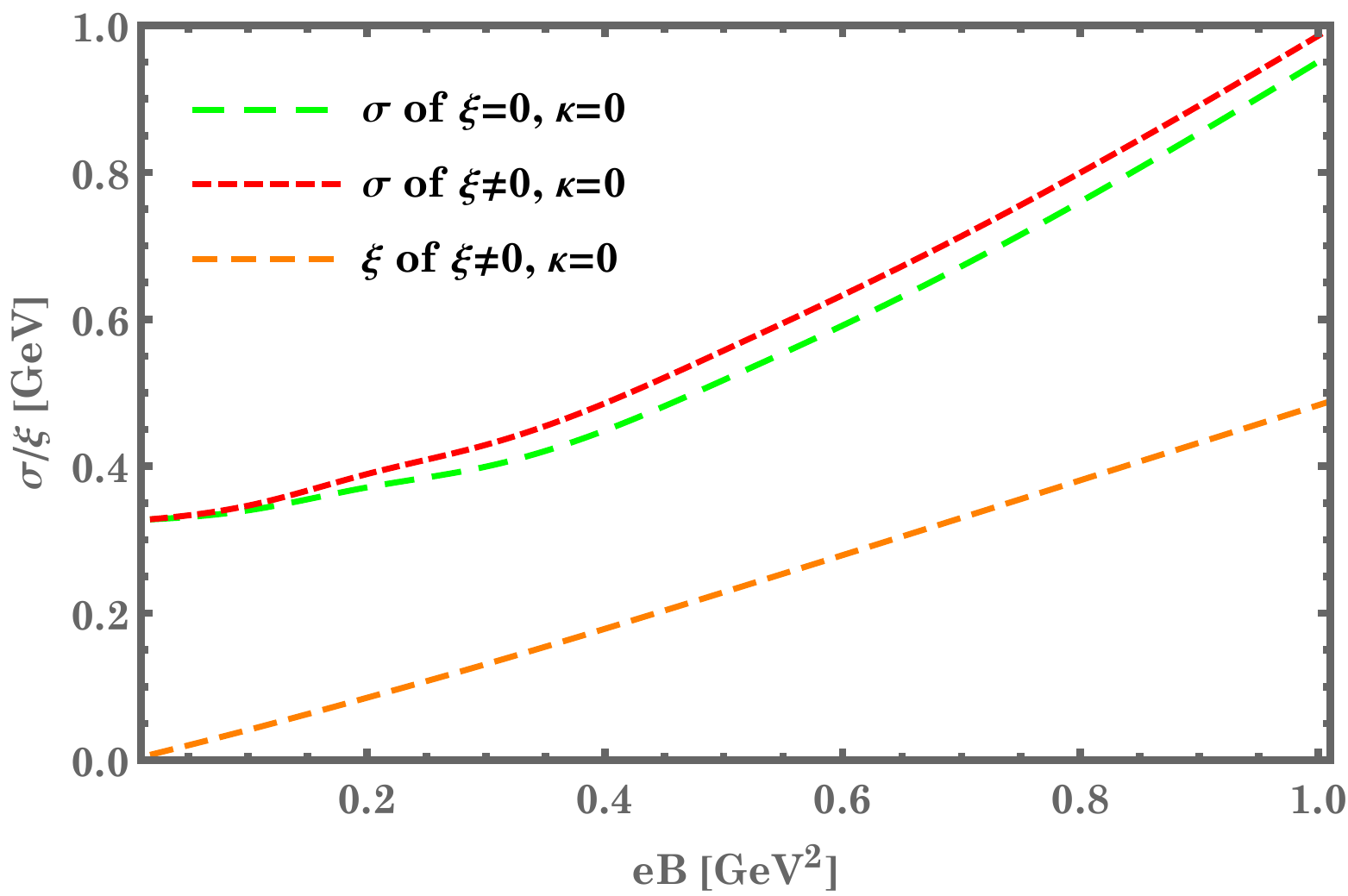}}
 	\subfigure[]{
 		\label{Fig1b}
 		\includegraphics[scale=0.55]{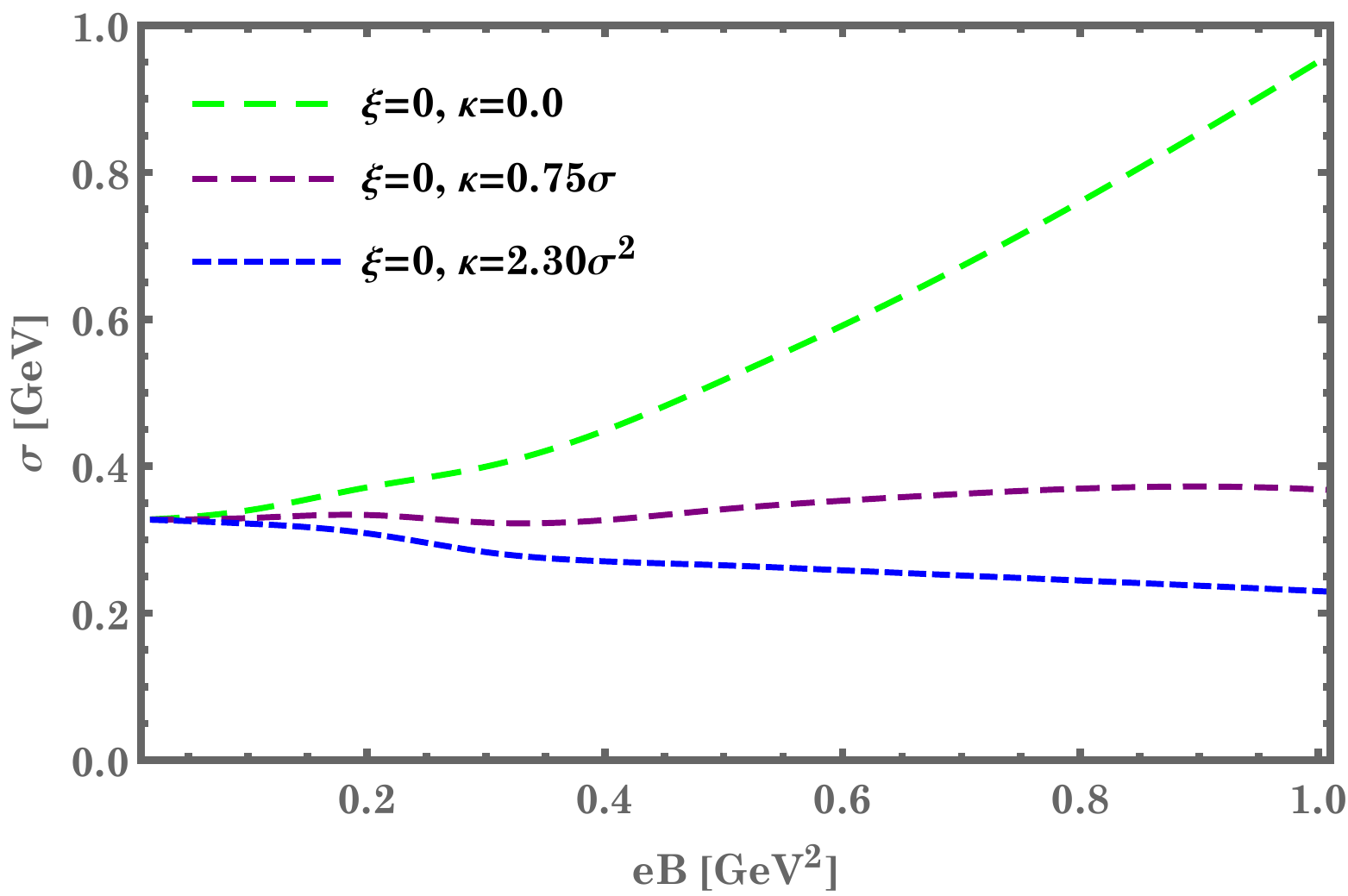}} 
 	\caption{Condensates as functions of magnetic field. (a) The chiral condensate $\sigma$ and TSP $\xi$ (TSP) as functions of $eB$. (b) The chiral condensate $\sigma$ as a function of $eB$ with different forms of $\kappa$ (AMM). }
 	\label{Fig1}
 \end{figure}

The full numerical calculation of Eq.(\ref{gapeq-xi-chiral}) for the chiral condensate and tensor-type spin polarization is shown in Fig.\ref{Fig1a}. It is found that the TSP $\xi$ increases linearly with the magnetic field $eB$, and the chiral condensate $\sigma$  increases slightly compared with the case of $\xi=0$, which confirms the analytic result at strong magnetic field, indicates that the TSP further enhances the magnetic catalysis. 
 
\textbf{Magnetic Inhibition in the vacuum and Inverse Magnetic Catalysis with AMM at finite temperature:}

By neglecting the current mass of quarks and applying a hard cut-off, Eq.(\ref{gapeq-AMM-chiral}) has form of :
\begin{equation}
	\frac{\pi^2}{G_SN_c}=\sum_{f} |q_{f}B|(1- 2 v\sigma |q_{f}|B)(1-v \sigma|q_{f}|B)\, \mathrm{arcsinh}\frac{\Lambda}{\sigma-v \sigma^{2}|q_{f}|B}.
\end{equation}
This equation could not be  solved analytically even under reasonable approximations, but the physical condition requires
\begin{equation}
{\sigma-v \sigma^{2}|q_{f}|B}>0 \quad  \text{and} \quad  \sigma>0, \quad \Longrightarrow \quad \sigma < \frac{1}{v |q_{f}|B},
\end{equation} 
which gives an upper bound on $\sigma$.

\begin{figure}[H]
	\centering 
		\includegraphics[scale=0.54]{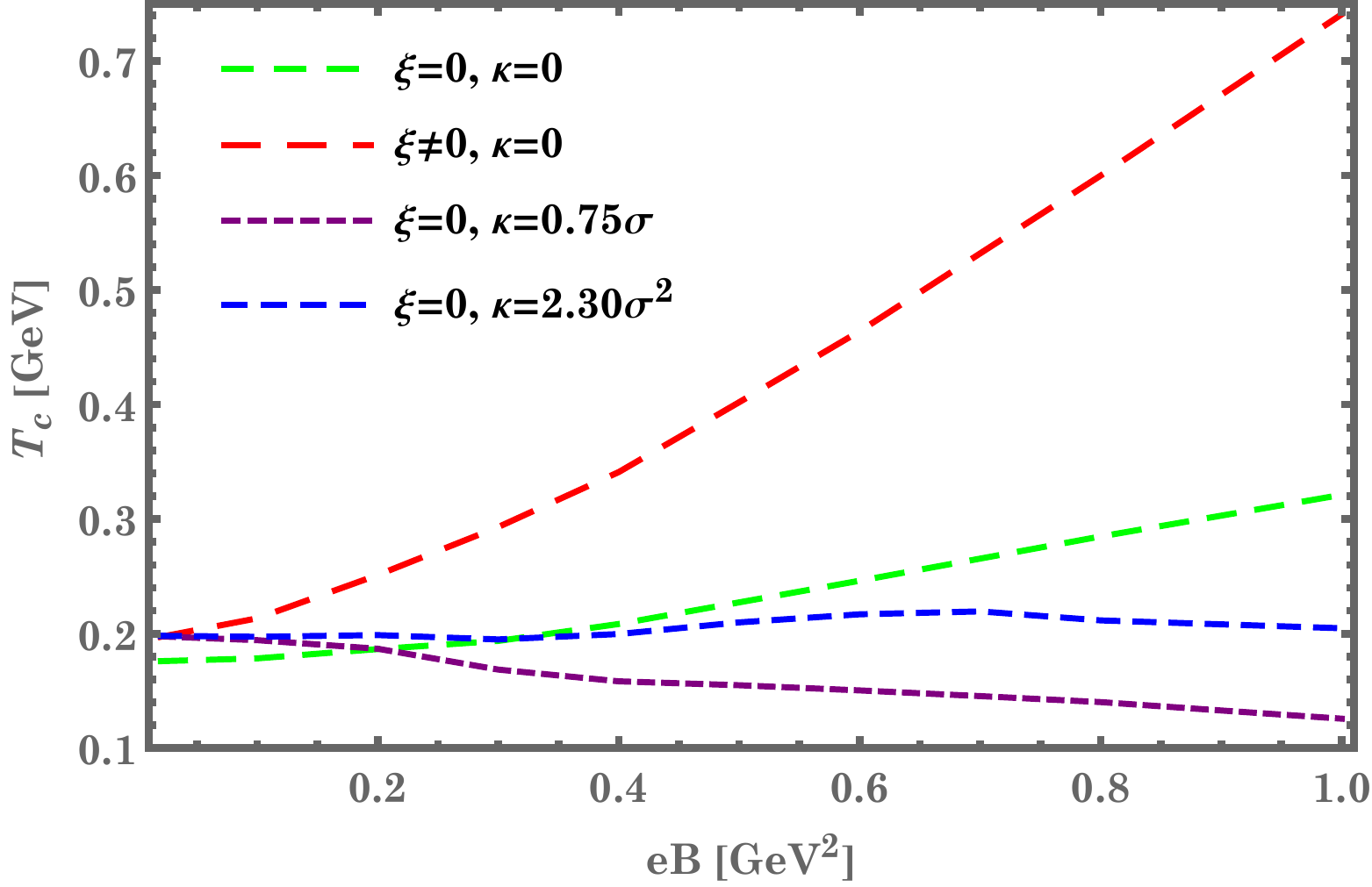}
	\caption{Critical temperature of chiral phase transition as function of $eB$. }
	\label{FigTcAMMIMC}
\end{figure}

The full numerical results of the chiral condensate Eq.(\ref{gapeq-AMM-chiral}) with different forms of AMM is shown in Fig.\ref{Fig1b}. It is clearly seen that the AMM of quarks reduces the chiral condensate in the vacuum, which can be called the  Magnetic Inhibition effect \cite{Fukushima:2012kc}. Obviously, this Magnetic Inhibition at finite temperature exhibits the IMC effect as shown in Fig.\ref{FigTcAMMIMC}.

\subsection{Magnetic Susceptibility}

Recently the lattice QCD calculation showed  that the magnetized QCD matter exhibits diamagnetism (negative susceptibility) at low temperature and paramagnetism (positive susceptibility) at high temperature~\cite{Bali:2012jv,Bali:2020bcn}. Therefore we calculate the magnetic susceptibility in the NJL model with TSP and AMM, respectively.

\par
\begin{figure}[H]
	\centering 
	\includegraphics[scale=0.55]{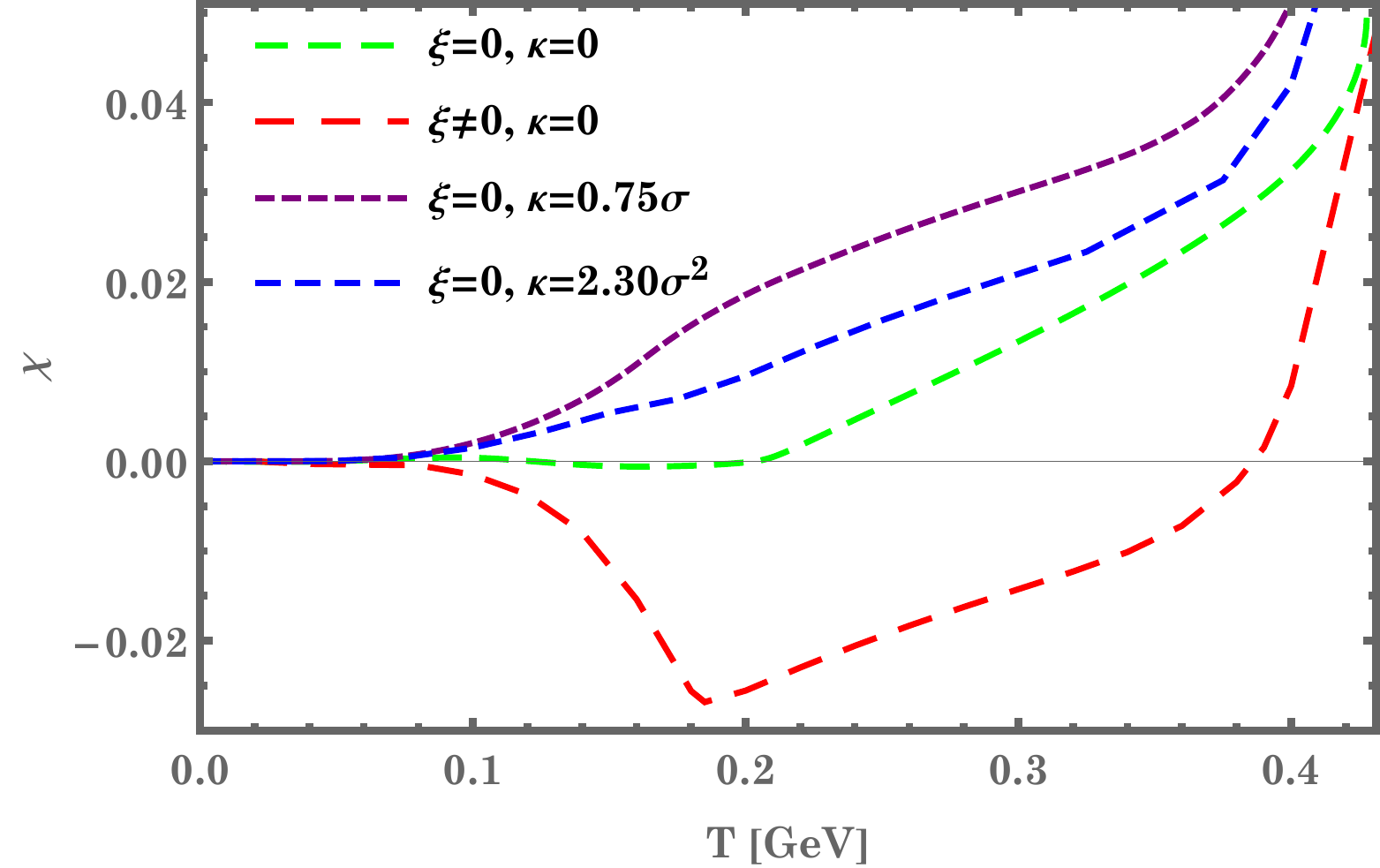}
	\caption{ Magnetic susceptibility $\chi$ as a function of the temperature $T$ with AMM $\kappa$ and TSP $\xi$, respectively. }
	\label{FigSus}
\end{figure}

 The  magnetic susceptibility is defined as:
\begin{equation}
\bar{\chi}(T)=-\frac{\partial^2 \Omega(T,eB)}{\partial (eB)^2}\Big|_{eB=0},
\end{equation}
in the lattice calculation, the renormalization scale choice fixes $ \bar{\chi}(0)=0 $ to eliminate unobservable vacuum magnetic susceptibility, so we naturally define a new magnetic susceptibility with a direct shift $\chi(T)=\bar{\chi}(T)-\bar{\chi}(0)$:
 \begin{equation}
 \chi(T)=\bar{\chi}(T)-\bar{\chi}(0)=\frac{\partial^2 \Delta(T,eB)}{\partial (eB)^2}\Big|_{eB=0},
 \end{equation}
with
\begin{equation}
	\Delta(T,eB)= \Omega(0,eB)-\Omega(T,eB)=N_c\sum_{f,n,s}\alpha_{n,s}\frac{|q_{f}B|}{2\pi}\int_{-\infty}^{+\infty} \frac{\mathrm{d}p_z}{2\pi} 2T \ln(1+\text{e}^{-\beta E_{f,n,s}}).
\end{equation}
 After some simple derivation we get:
 \begin{equation}
 	\frac{\partial^2 \Delta(T,eB)}{\partial (eB)^2}=2 N_c\sum_{f,n,s}\frac{|q_{f}|}{2\pi}\int_{-\infty}^{+\infty} \frac{\mathrm{d}p_z}{2\pi}\  ({1+\mathrm{e}^{\beta E_{f,n,s}}})^{-1}\left[ (1+\text{e}^{-\beta E_{f,n,s}})^{-1}\beta|eB|\left(\frac{\partial E_{f,n,s}}{\partial eB}\right)^{2}-|eB|\frac{\partial^2 E_{f,n,s}}{\partial (eB)^2}-2\frac{\partial E_{f,n,s}}{\partial eB}
 	\right].
 \end{equation}
However, the above expression does not has a good analytical definition for $ eB=0 $, though the summation over Landau levels is convergent for arbitrary $|eB|>0$. Thus, we determine the magnetic susceptibility in the vicinity of $eB=0$ numerically: 
 \begin{equation}
 	\chi(T)=\frac{\Delta(T,2\epsilon)-2\Delta(T,\epsilon)+\Delta(T,0)}{\epsilon^{2}},
 \end{equation}
we choose $\epsilon=0.01 $ and sum $5000$ Landau levels to ensure stable numerical result as shown in Fig.\ref{FigSus}. It is observed that 1) in the case without AMM and TSP, i.e, $\kappa=0, \xi=0$, the  magnetic susceptibility is zero in the temperature region of $T<T_c$, and becomes positive at $T>T_c$; 2) in the case with non-zero  AMM: $\kappa=0,75 \sigma$ or $\kappa=2.5 \sigma^2$, the magnetic susceptibility is always positive and increases with the temperature $T$; 3) in the case of TSP $\xi$, the magnetic susceptibility is negative at low temperature and positive at high temperature. 

From above numerical results, we can draw the conclusion that the AMM induces paramagnetism while the TSP induces the diamagnetism at low temperature.

\begin{figure}[H]
 \subfigure[]{
 		\label{tensorSp1}
 		\includegraphics[scale=0.55]{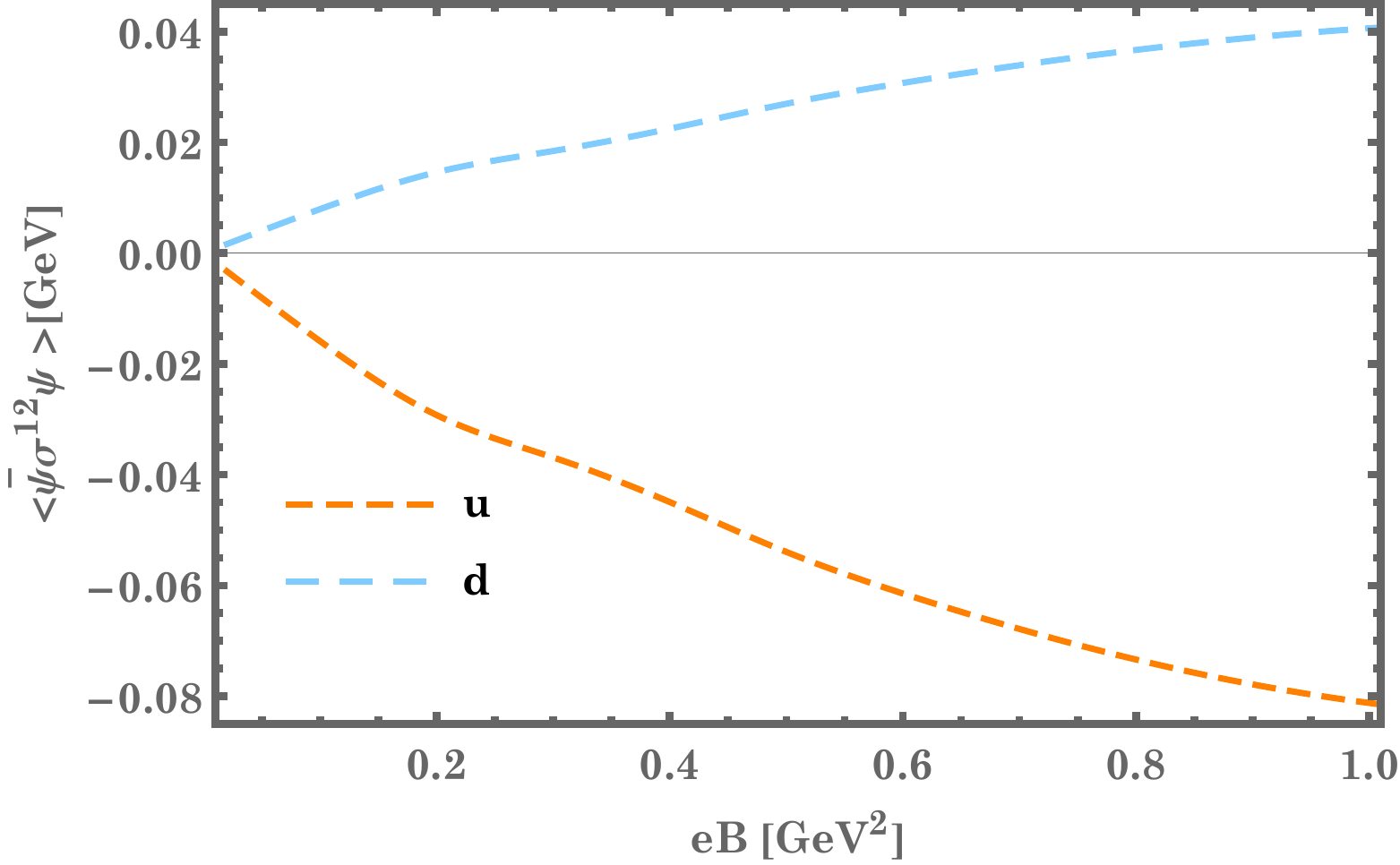}}
 	\subfigure[]{
 		\label{tensorSp2}
 		\includegraphics[scale=0.55]{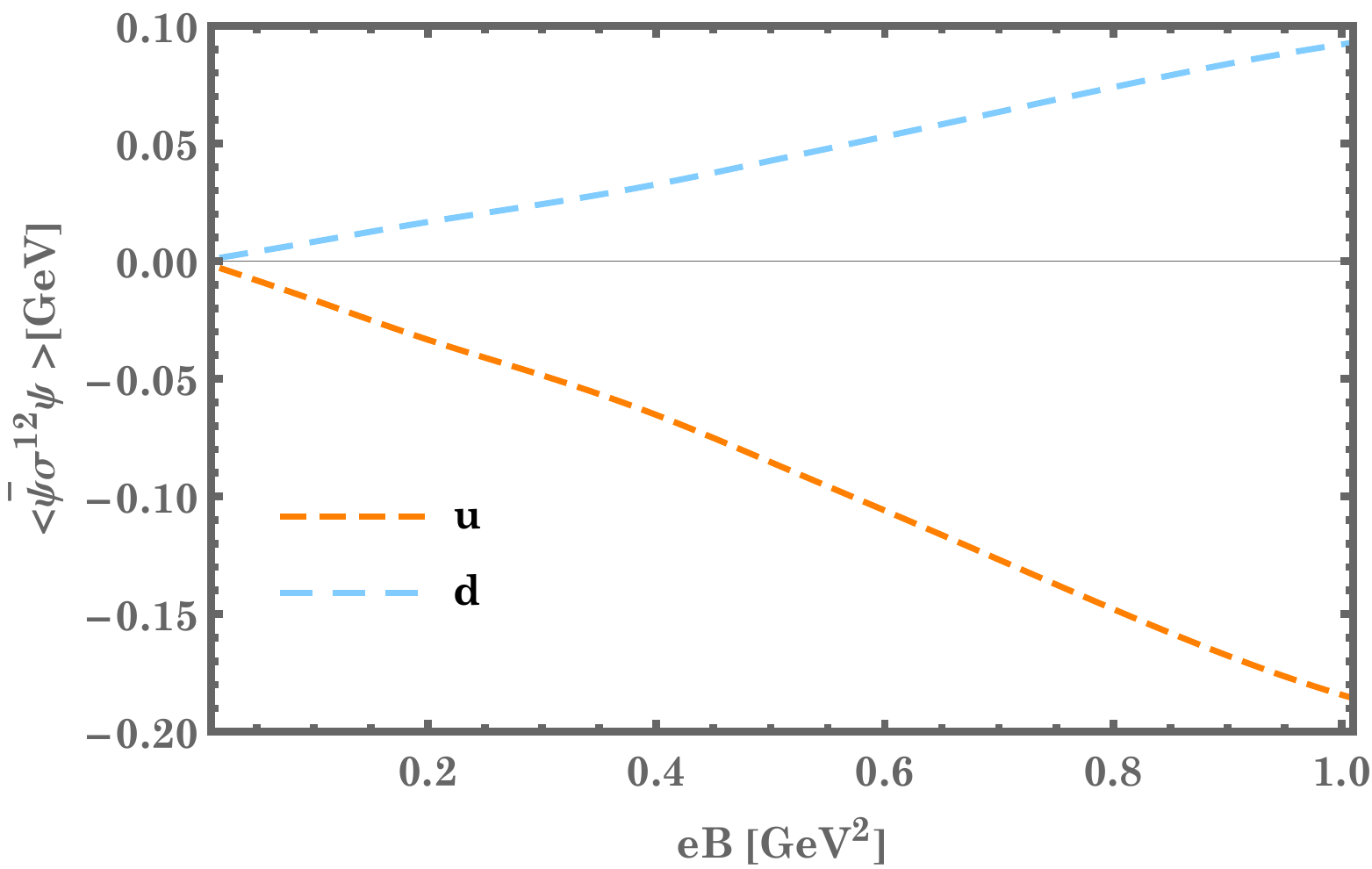}}
 	\caption{Spin polarization of u/d quarks change with magnetic field.  (a) Condensate $\left< \bar{\psi}\sigma^{12}\psi \right>$ of u and d quarks with AMM $\kappa=2.30\,\sigma^{2}$. (b) Condensate  $\left< \bar{\psi}\sigma^{12}\psi \right>$ of u and d quarks with AMM $\kappa=0.75\,\sigma$.}
 	\label{tensorSp}
 \end{figure}

\textbf{ Tensor spin condensate for different flavors:}  Reference \cite{Bali:2020bcn} considered the spin condensate for different flavors in lattice, which is related to the magnetic susceptibility.
The tensor spin condensate of u/d quark can be extracted from the AMM term through
\begin{equation}
\langle\bar{\psi}\sigma^{\mu\nu}\psi\rangle=-\kappa q_{f}F^{\mu\nu},
\end{equation} 
whose non-zero components are $\langle\bar{\psi}\sigma^{12}\psi\rangle,\langle\bar{\psi}\sigma^{21}\psi\rangle$ in uniform magnetic along $z$ direction shown in Fig.\ref{tensorSp}.
 
It is found that the tensor spin condensate of u(d) quark decreases(increases) linearly with the magnetic field when taking $\kappa\varpropto \sigma$,  which is similar as found in Ref.\cite{Bali:2020bcn}. It should be pointed out that the tensor spin condensate extracted from AMM in Fig.(\ref{tensorSp}) is not the same quantity as the tensor spin polarization $\xi$, which is charge independent.

\subsection{Neutral and Charged Pion Mass Spectra}
\label{sec-meson}
In this section, we investigate the neutral and charged pion mass with the TSP condensate and AMM term, respectively. In the NJL model, mesons are constructed as $q\bar q $ bound states or resonances obtained from the quark-antiquark scattering amplitudes through four-fermion interaction~\cite{Klevansky:1992qe}. The mass of the meson can be obtained using Random Phase Approximation(RPA) in the leading order of $1/N_c$ expansion \cite{tHooft:1974pnl,Witten:1979kh}. Following the procedure from \cite{Liu:2014uwa,Liu:2015pna,Miransky:2015ava}, the quark propagator in the Landau levels takes the following form
\begin{equation}
S(x,y)=\text{e}^{i\Phi_f(x,y)}\int \frac{d^4 p}{(2\pi)^4}\text{e}^{-ip(x-y)}\widetilde{S}(p),
\end{equation}
where the Schwinger phase $\Phi_f(x,y)=q_f (x^1+y^1)(x^2-y^2)/2$ breaks the translation invariant while $\widetilde{S}(q)$ is translation-invariant
\begin{equation}
	\label{eq:propagator}
	\widetilde{S}(p)=i e^{-p_{\perp}^{2} l^{2}} \sum_{n=0}^{\infty}(-1)^{n}  \frac{D_{n}(p)}{\mathcal{M}-2 n|q_f e B|}.
\end{equation}
The denominator part has the form of
\begin{equation}
\frac{1}{\mathcal{M}-2 n|e B|}=\frac{ \left(p^0\right)^2-\left(p^3\right)^2-m^2+\tilde{\mu}^2-2n|q_f eB|- 2p^3 \tilde{\mu} \gamma^0 \gamma^5 + 2p^0 \tilde{\mu}  \gamma^3 \gamma^5
}
{\left[\left(p^0\right)^2-\left(p^3\right)^2-m^2-\tilde{\mu}^2-2n|q_f eB| \right] ^2-8n|q_f eB|\tilde{\mu}^2-4 m^2 \tilde{\mu}^2
}, 
\end{equation}
and the numerator takes the form of
\begin{equation}
	D_{n}(p) =  2\left(p^0 \gamma^{0} - p^{3} \gamma^{3}+M-i \tilde{\mu} \gamma^1 \gamma^2 \right)\left[\mathcal{P}_{+} L_{n}\left(2 p_{\perp}^{2} l^{2}\right)-\mathcal{P}_{-} L_{n-1}\left(2 p_{\perp}^{2} l^{2}\right)\right]+2  \left( p^1 \gamma^1 + p^2 \gamma^2  \right) L_{n-1}^{1}\left(2 p_{\perp}^{2} l^{2}\right),
\end{equation}
where we introduced the magnetic length $l=1/\sqrt{|q_f B|}$ and transverse momentum $ p_\perp=(0,p^1,p^2,0)$, $L_n^{\alpha}$ are the generalized Laguerrel polynomials with $L_n=L_n^0 , L_{-1}^{\alpha}=0 $, the $\tilde{\mu}$ means tensor condensate here equating to TSP $\xi$ or AMM $\kappa$ . As is seen, The spin projection operators 
\begin{equation}
	\mathcal{P}_{\pm}=\frac{1}{2}\left[1 \pm i \gamma^{1} \gamma^{2} \operatorname{sign}( q_f B)\right]
\end{equation}
emerge in the propagator creating the spin subspaces with opposite spin projection under SP operator $ i \tilde{\mu} \gamma^1 \gamma^2 $, which causes the split of degenerate energy levels. 
\par Now, we are ready to study the effect of SP on meson mass spectrum. Under RPA approximation, we sum all chain-typed Feynman diagrams constituted by one quark loop, then obtain effective interaction between a pair of quarks ($ud$ pair or $uu,dd$ pair) \cite{Klevansky:1992qe}:
\begin{equation}
	D_{\pi}(q^2)=\frac{2G_{S}}{1-2G_{S}\Pi_{\pi}(q^2)},
\end{equation}
which is equivalent to the pion meson as intermediate boson, thus, the pole of propagator $D_{\pi}(q^2)$ obviously gives the pion mass by gap equation:
\begin{equation}
1-2 G_S \Pi_{\pi}(q^2=m_{\pi}^2)=0,
\end{equation}
where the one-loop polarization function of pion $\Pi_{\pi}$ by one quark loop is \cite{Klevansky:1992qe}:
\begin{eqnarray}
	\label{pola_pi}
	\Pi_{\pi}(q^2)  =  i \int \frac{d^4k}{(2\pi)^4} \text{Tr}[\tau^a i\gamma^5 \widetilde{S}(k)\tau^b i\gamma^5 \widetilde{S}(p)],
\end{eqnarray}
with momentum conservation condition $q=k-p$. For neutral pion, the Pauli matrices are chosen $\tau^a=\tau^3,\tau^b=\tau^3$  while for charged pion $\tau^a=\tau^{\pm},\tau^b=\tau^{\mp},\tau^{\pm}=(\tau^1\mp i\tau^{2})/\sqrt{2}$. In view of the complexity of propagator, we also calculate the pion mass numerically.

\begin{figure}[H]
	\centering 
	\subfigure[]{
		\label{Fig3a}
		\includegraphics[scale=0.55]{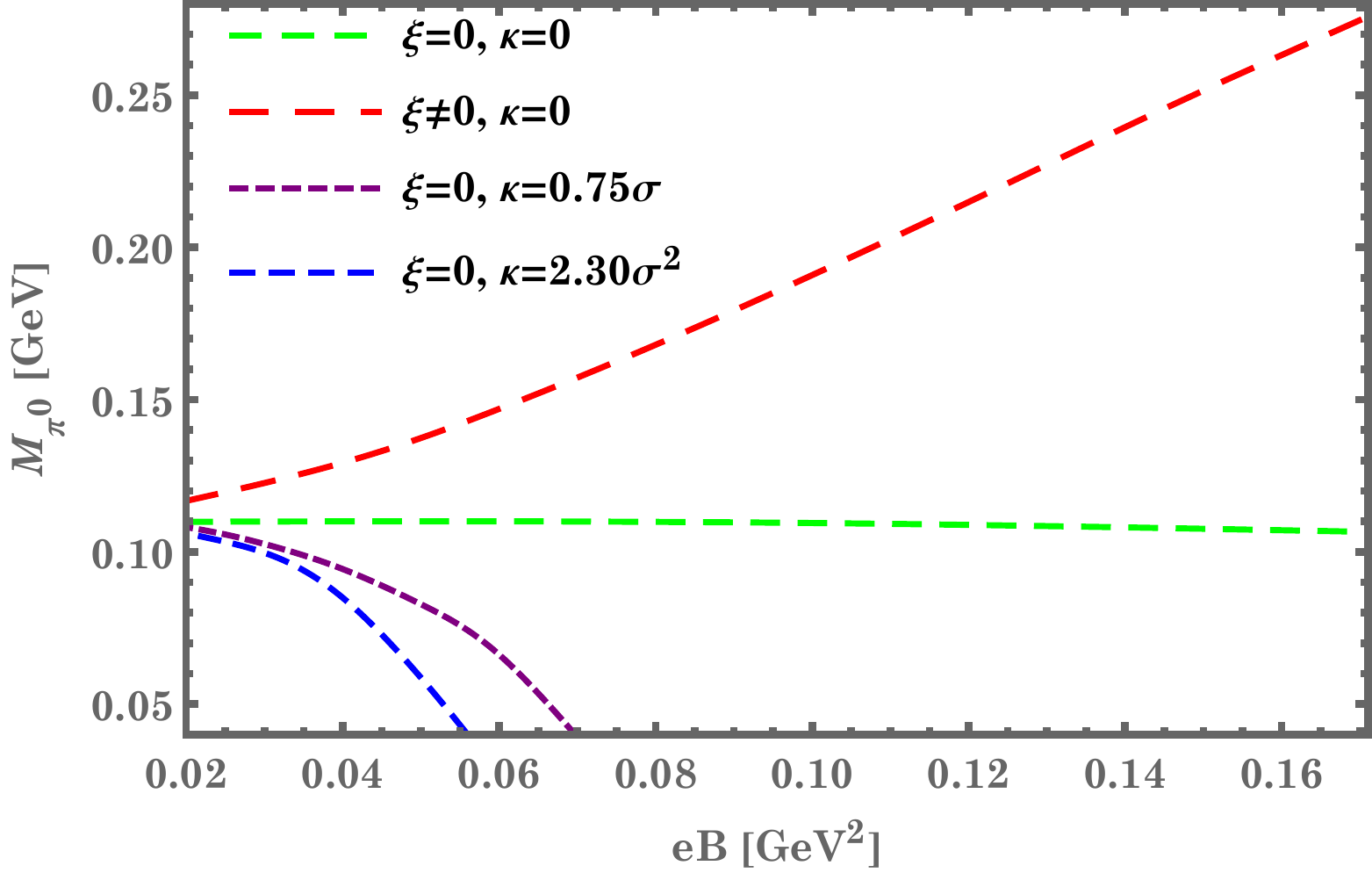}}
	\subfigure[]{
		\label{Fig3b}
		\includegraphics[scale=0.55]{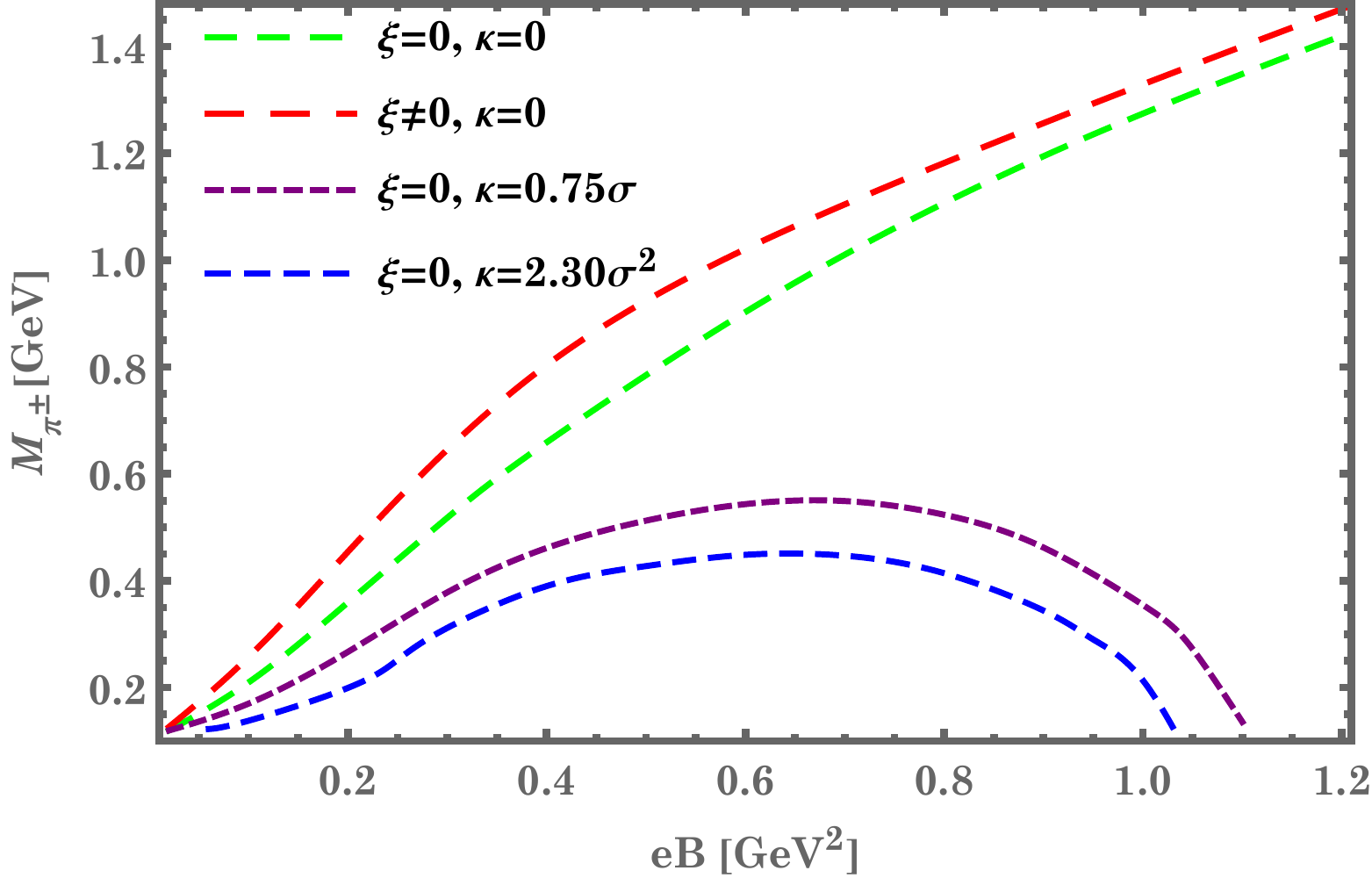}} 	
	\caption{Pion mass as a function of the magnetic field. (a) Neutral pion mass $m_{\pi^0}$ as a function of $eB$ with TSP and magnetic dependent AMM. (b) Charged pion mass $m_{\pi^\pm}$ as a function of $eB$ with TSP and magnetic dependent AMM. }
	\label{Figpion}
\end{figure}

In Fig.\ref{Figpion}, we show the neutral and charged pion mass change with the magnetic field in the case of considering AMM contribution $\kappa$ and tensor-type spin polarization $\xi$, respectively. It is observed that with TSP $\xi$, both the neutral pion mass and charged pion mass increases with the magnetic field, which might be related to the magnetic catalysis effect of TSP $\xi$. And in the case of considering magnetic dependent AMM contribution, it is found that neutral pion mass decreases with the magnetic field, and charged pion mass shows non-monotonic behavior, i.e., firstly increases then decreases with the magnetic field. It is worth pointing out that, we considered the case with constant AMM $\kappa$ in \cite{Xu:2020yag}, but the charged pion mass does not show non-monotonic behavior. The behavior of neutral and charged pion mass in case with magnetic dependent AMM contribution is qualitatively in agreement with lattice result in \cite{Ding:2020jui}.

\section{Discussion and Conclusion}
\label{sec-con}
In this work, we thoroughly investigate the effect from the TSP and magnetic dependent AMM on the magnetized vacuum, magnetic susceptibility and pion mass.

It is observed that with the magnetic dependent AMM $\kappa\varpropto \sigma$
or $\kappa\varpropto \sigma^2$, we have the magnetic inhibition of the chiral condensate in the vacuum and inverse magnetic catalysis around the critical temperature, and related to this property, we also found that the neutral pion mass decreases with the magnetic field and the charged pion mass firstly increases then decreases with  the magnetic field. However, the diamagnetism or negative magnetic susceptibility at low temperature can not be explained in this case.

With TSP $\xi$,  one can explain the diamagnetism, i.e, the negative magnetic susceptibility at low temperature. This can be understood because TSP with parallel spin pairing of quark anti-quark does not exhibit net MM. Therefore, the total net MM of the system with TSP condensate reduces comparing with the system with chiral condensate carrying net MM. However, the TSP further enhances the magnetic catalysis and we cannot get the IMC effect around the critical temperature, and both the neutral and charged pion mass increase with the magnetic field, these behaviors are opposite to lattice results in \cite{Ding:2020jui}.

Therefore, our results suggest that a magnetic dependent AMM contribution is reasonable to be taken into account in the magnetized quark system. The diamagnetism property might be induced by another mechanism related to orbital angular momentum. It is known that in the classical quantum statistics, fermion system in weak magnetic field manifests famous Landau diamagnetism from the orbital motion contribution and Pauli paramagnetism from spin coupled magnetic moment at low and high temperatures respectively. We leave this for future studies.
\begin{acknowledgements}
	
This work is supported in part by the National Natural Science Foundation of China (NSFC)  Grant  Nos. 11735007, 11725523, and Chinese Academy of Sciences under Grant Nos.XDB34000000 and XDPB15, the start-up funding from University of Chinese Academy of Sciences(UCAS), and the Fundamental Research Funds for the Central Universities.
\end{acknowledgements}

\end{document}